\documentclass[aps,pra,groupedaddress,twocolumn]{revtex4}
\usepackage{graphicx,amsmath,amssymb}

\begin{document}

\title{Dissipative solitons and vortices in polariton superfluids}
\author{Elena A. Ostrovskaya$^{1}$, Jasur Abdullaev$^{1}$, Anton S. Desyatnikov$^{1}$, Michael D. Fraser$^{2}$, and Yuri S. Kivshar$^{1}$}
\affiliation{$^{1}$Nonlinear Physics Centre, Research School of Physics and Engineering, The Australian National University, Canberra, ACT 0200, Australia}
\affiliation{$^{2}$National Institute of Informatics, Tokyo, Japan}

\date{\today}

\begin{abstract}
We examine spatial localisation and dynamical stability of Bose-Einstein condensates of exciton-polaritons in microcavities under the condition of off-resonant spatially inhomogeneous optical pumping both with and without a harmonic trapping potential. We employ the open-dissipative Gross-Pitaevskii model for describing an incoherently pumped polariton condensate coupled to an exciton reservoir,  and reveal that spatial localisation of the steady-state condensate occurs due to the effective self-trapping created by the polariton flows supported by the spatially inhomogeneous pump, regardless of the presence of the external potential. A ground state of the polariton condensate with repulsive interactions between the quasiparticles represents a dynamically stable bright dissipative soliton. We also investigate the conditions for sustaining spatially localised structures with non-zero angular momentum in the form of single-charge vortices. 
\end{abstract}

%\pacs{03.75.Lm, 03.75.Kk, 05.60.-k}

\maketitle

\section{Introduction}
Since the first observation of spontaneous exciton-polariton condensation in a microcavity \cite{BEC06}, significant effort in this vigorous field has been directed towards understanding the properties of superfluid flow in the  novel non-equilibrium polariton superfluid \cite{pvortex08,superflow, Vortex11,vortex11_2}.  Recently, considerable attention was drawn to the controlled creation of spatially localised collective excitations of the exciton-polariton Bose-Einstein condensate (BEC). Most notably, moving bright solitons have been predicted \cite{skryabin} and observed in a pioneering proof-of-principle experiment \cite{sich}. The properties of polariton solitons, superior to those of optical solitons in semiconductor cavity lasers, namely  the picosecond response times and large nonlinearities,  suggest that polariton BEC can offer novel functionalities for information-processing devices \cite{rosanov_nature}. The bright solitons observed in \cite{sich} require a resonant excitation regime at nonzero in-plane momentum to make use of the dispersion properties of the lower-polariton branch that allow the polaritons to have a negative effective mass. In this regime, strong repulsive interactions between the quasiparticles can be balanced by the effective dispersion to achieve localisation. Similar mechanism for spatial localisation via effective mass management is well explored for atomic BECs in optically induced band-gap structures \cite{gap}. In the absence of the spectral gap, however, localisation by means of effective mass management is only possible in one spatial dimension. The question whether localisation of polariton BEC with repulsive interactions can occur in  the regime of spontaneous condensation at zero in-plane momentum (positive effective mass) and non-resonant excitation, remains open.

In this paper we examine, both analytically and numerically, formation and dynamical stability of a steady-state BEC of exciton-polaritons in microcavities in the presence of a spatially localised (Gaussian) optical pump. By employing the open-dissipative Gross-Pitaevskii model \cite{Wouters07} describing an incoherently pumped BEC coupled to the exciton reservoir and successfully used in theoretical description of a number of significant experiments (see, e.g., \cite{pvortex08, Vortex11}), we analyse the mechanisms for creating and sustaining two-dimensional spatially localised structures, such as dissipative solitons and vortices. In addition to the trap-free case, we analyse the structure of the localised states with the addition of a harmonic external potential that can be created, e.g., by engineered stress of the microcavity \cite{Balili07}. In the latter case, the localisation of the repulsive polariton BEC is due to the harmonic confinement, as is the case with the analogous trapped atomic BEC with a positive scattering length.

\begin{figure}[here]
\includegraphics[width=8.5 cm]{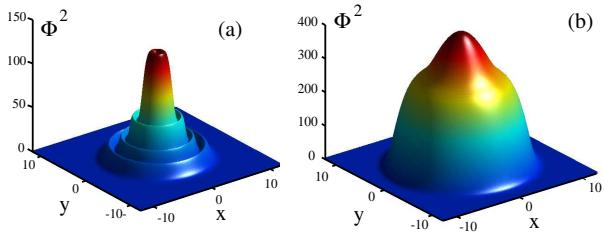} 
\caption{Typical steady state density of a harmonically trapped polariton condensate in the regime of (a) narrow and (b) wide pump (see text).}
\label{3d}
\end{figure}

We show, analytically and numerically, that, even in the absence of external potentials localisation of the steady-state polariton BEC occurs  due to the effective self-trapping created by the polariton flows due to the spatially inhomogeneous off-resonant optical pumping. Spatial localisation of the condensate occurs due to the internal balance of superfluid density flows which is identical to that responsible for supporting dissipative "anti-solitons" in optical systems. Such localisation of continuously self-defocusing solitons occurs despite the repulsive interactions between the quasiparticles, and is analogous to optical Ògain-guidingÓ effect \cite{siegman,1d_kartashov_ol} in its reliance on continuous pumping. Thus a {\em ground state of the polariton BEC} represents a {\em dissipative soliton} which is spatially localised and dynamically stable, regardless of the presence of the external potential. A harmonic trapping potential dramatically modifies the structure of the steady state due to the competition with the spatially inhomogeneous pump (see Fig. \ref{3d}).

A single vortex created in a localised steady state BEC by phase imprinting is similarly supported by the continuous inhomogeneous pump as a spatially localised {\em dissipative vortex soliton}. Akin to a vortex line in a trapped atomic BEC with a positive scattering length it is dynamically unstable \cite{vortex_atom}.  In the absence of potential, a vortex line spirals out of the polariton condensate and the condensate restores to its ground steady state. We show that the addition of fabricated (i.e. non-rotating) harmonic potential modifies the threshold of the optical pumping required to sustain a steady state with an angular momentum and leads to the possibility of a long-term survival of the vortex. 

\section{Model}
We consider a spontaneously formed exciton-polariton BEC (${\vec k}=0$, lower-polariton branch) under the continuous-wave non-resonant excitation. The model first suggested in \cite{Wouters07} consists of a mean-field equation for the polariton condensate wavefunction and a rate equation for the inhomogeneous density of the exciton reservoir. We write it here in the form used in \cite{Vortex11}:  
\begin{eqnarray} 
i\hbar\frac{\partial \Psi}{\partial t}&=&\left[-\frac{\hbar^2}{2m}\nabla_\perp^2+V(\vec{r},t)+ i\frac{\hbar}{2}(Rn_R-\gamma_c)\right]\Psi,  \nonumber \\  
\frac{\partial n_R}{\partial t}&=&-(\gamma_R+R|\Psi|^2)n_R(\vec{r},t)+P(\vec{r}), \label{model} 
\end{eqnarray}
where $V(\vec{r},t)=U(\vec{r})+g_c|\Psi|^2+g_Rn_R(\vec{r},t)$. Here $\Psi$ is the condensate wavefunction, $n_R$ is the exciton reservoir density, $P(\vec{r})$ is the inhomogeneous optical pump, and $U(\vec{r})$ is the external potential. The critical parameters defining the condensate dynamics are the loss rates of the polaritons $\gamma_{c}$ and reservoir excitons $\gamma_{R}$, the stimulated scattering rate $R$, condensate coupling to the reservoir $g_R$, and the coefficient $g_c$ quantifying the nonlinear interaction of polaritons. In what follows, we will consider a radially symmetric external trapping potential $U(r)=U^2_0r^2$, where $U_0=0$ in a trap-free case.

The model can be re-written in dimensionless form by using the following characteristic scaling units of time, energy, and length:
\begin{equation}
T=1/\gamma_c, \quad E=\hbar \gamma_c,  \quad L=\sqrt{\frac{\hbar}{2m_{LP}\gamma_c}},
\label{scale}
\end{equation}
where $m_{LP}$ is the lower-polariton effective mass. Here, we choose the parameters close to those of the experimental setup of \cite{Vortex11}, with $m_{LP}=10^{-4}m_e$, $g_c=6\times 10^{-3}$ $meV$$\mu m^2$, $g_R=2g_c$, $\gamma_c=0.33$ $ps^{-1}$, $\gamma_r=1.5 \gamma_c$, and $R=0.01$ $\mu m^2$ $ps^{-1}$. In what follows we use the dimensionless (normalised) variables and parameters.

We recall that, in the spatially homogeneous case, one can estimate the threshold pumping power at which the condensate appears in the system, as described in \cite{Wouters07}. This threshold is approximately determined by the values of parameters for which stimulated gain equals loss:
\begin{equation}
P_{th}\approx \frac{\gamma_R\gamma_c}{R}.
\label{th}
\end{equation}
The ratio of the polariton $n_c$ to exciton reservoir $n_R$ densities in the steady homogeneous state can then be defined as:
\begin{equation}
\frac{n_c}{n_R}=\frac{\gamma_c}{\gamma_R}(\bar{P}-1),
\label{ratio}
\end{equation}
where normalised pump is defined as $\bar{P}=P/P_{th}\geq 1$. In general, even in a spatially inhomogeneous case, the typical dynamical behaviour of the model Eqs. (\ref {model}) displays rapid transition to the steady state of the polariton BEC for $\bar P>1$. 

Below we consider the properties of the steady state supported by an inhomogeneous (Gaussian) optical pump,  $P(r)=P_0\exp(-r^2/\sigma^2)$, with or without an additional trapping potential. The spot size of the optical pump is considered to be small compared to the size of the sample, so that the local density approximation \cite{Wouters08} is not applicable.

\section{Steady state of the condensate}
\subsection{Theory}
For a general CW pump, the condensate wavefunction in the steady state can be looked for in the following general form: $\Psi=\psi(x,y)\exp(-i\mu t)$, where $\mu$ is the chemical potential. If both the external potential and the pump are radially symmetric, we can re-write the model system in polar coordinates $(x,y) \to (r,\theta)$ and reduce the problem to finding radially symmetric steady states $\psi(x,y)=\psi(r)\exp(im\theta)$, where $m$ is the phase winding number \cite{anton} (topological charge of a vortex), and the ground state corresponds to $m=0$.

The steady state wavefunction $\psi(r)$ obeys the following stationary equation:
\begin{equation}
\nabla_r^2\psi-2V(r)\psi-i(Rn^0_R-\gamma_c)\psi+2\mu \psi=0,
\label{stat}
\end{equation}
here $\nabla_r^2=\partial^2/\partial r^2+(1/r)\partial/\partial r-m^2/r^2$, $V(r)=g_c|\psi|^2+g_Rn^0_R+U(r)$, and the steady state reservoir density is found from Eqs.  (\ref{model}) as:
\begin{equation}
n^0_R=\frac{P(r)}{\gamma_R+R|\psi|^2}.
\end{equation}
All physical parameters in Eq. (\ref{stat}) are dimensionless, and under the adopted scaling (\ref{scale}), $\gamma_c=1$. However, we formally retain this parameter in subsequent formulas.

The stationary wave function can be further separated into the real amplitude and phase: $\psi(r)=\Phi(r)\exp[i\phi(r)]$. Following the analysis suggested in \cite{rosanov} for the generalised complex Ginzburg-Landau equation, we can obtain asymptotic behaviour of amplitude and phase of the condensate wavefunction. In the absence of a trapping potential, $U(r)=0$, the requirement of spatial localisation, $P(r)\to 0$, $|\psi(r)|\to 0$, imposed on both the pump and the condensate wave function at $r\to \infty$ leads to the following asymptotic behaviour of the amplitude and the phase at large $r$:
\begin{equation}
\Phi(r)\sim A\exp(-p_-r), \quad \phi(r) \sim B+p_+r
\label{infty}
\end{equation}
where $p_{\pm}=\left[ (\mu^2+\gamma^2/4)^{1/2}\pm \mu\right]^{1/2}$, and $A,B$ are real constants. We note that, at large $r$, and irrespective of the vorticity $m$, the condensate decays slower than the pump field. 

Conversely, at $r\to 0$, the asymptotic behaviour is found to be:
 \begin{equation}
 \Phi(r)\sim C r^{|m|}\exp(-q_-r^2), \quad \phi(r) \sim q_+r^2,
 \label{zero}
 \end{equation}
where $q_-=(1/2)(1+|m|)^{-1}[V(0)-\mu]$ and $q_+=(1/4)(1+|m|)^{-1}[Rn^0_R(0)-\gamma_c]$, and for the ground state $C=\Phi(0)\equiv\Phi_0$. Note that for $V(0)<\mu$ the second derivative of the amplitude changes sign at $r=0$, i.e. the condensate density displays a central "dip" in its spatial profile, as seen in Fig. \ref{3d}(a).

The presence of the harmonic trap $U(r)=U^2_0r^2$ dramatically modifies the asymptotic behaviour of the condensate wavefunction for large $r$ \cite{edwards95}, so that:
\begin{eqnarray}
 \Phi(r)\sim A \exp[-U^2_0r^2+(\mu-1/2){\rm ln}(2U_0r)], \label{inftytrap} \\ 
 \quad \phi(r) \sim B+\gamma_c {\rm ln} (2 U_0 r). \nonumber
 \end{eqnarray}
 The asymptotic behaviour of a trapped polariton BEC at $r\to 0$ is described by Eq. (\ref{zero}).

We note that for a spatially localised (e.g., Gaussian) pump, $P(r)$, the spatial localisation of the condensate in the absence of an added trapping potential is counterintuitive. Indeed, the effective potential $V(r)$ formed by the repulsive nonlinearity (due to polariton scattering) and interaction with the exciton reservoir is {\em anti-trapping}. Therefore the balance of nonlinearity and dispersion responsible for nonlinear localisation in conservative condensate systems cannot be achieved. Physically, the existence of the spatially localised steady state of the polariton condensate can be understood by examining the internal flows of the dissipative polariton superfluid, analogous to the Pointing vector flows in dissipative optical systems \cite{rosanov,akhmediev}. Indeed, by defining the superfluid current density (flux) in the standard way $\vec{ j}={\rm Im}(\Psi^*\nabla \Psi)$, we can re-cast Eq. (\ref{stat}) in the form of coupled equations for the amplitude of the condensate wave function, $\Phi$ and radial component of the stationary flux $J=j_r=n_c v_r$, where $n_c=\Phi^2(r)$ is the condensate density, and $v_r$ is the radial component of the flow velocity $\vec v=(d\phi/dr)\vec{e}_r+(m/r)\vec{e}_\theta$. The equations for these variables take the following form:
\begin{eqnarray}
\frac{1}{r}\frac{d}{d r}(rJ)-(Rn^0_R-\gamma_c)\Phi^2=0, \label{fluid} \\ 
\nabla_r^2\Phi-2\left[V(r)-\mu\right]\Phi-V_J\Phi=0, \nonumber
\end{eqnarray}
where $V_J=J^2\Phi^{-4}=(d\phi/dr)^2$.
The first equation of the system is the continuity equation for the stationary flow with source and sink. The first term in this equation is simply the divergence of the flux, $D=\nabla \cdot \vec j$, which serves as a local measure of gain ($D>0$) or loss ($D<0$). The steady state exists if the generation of the polariton superfluid via continuous pumping is balanced by its dissipation. This condition can be formulated in terms of the flux as follows \cite{akhmediev}:
\begin{equation}
\int^\infty_0 D(r) r dr=0.
\label{D}
\end{equation}
Remarkably, this condition can be satisfied {\em regardless of the sign of the nonlinear interactions} in the systems, i.e. for both repulsive (as in the case of polaritons) and attractive nonlinearities. Nonlinear eigenstates for a very similar dynamical 1D system with linear loss and spatially localised gain and without additional external potential were found in \cite{1d_kartashov_ol}. The key feature of our system is similar: for the given parameters of gain and loss, the chemical potential, $\mu$, is unique. 

The second equation in (\ref{fluid}) highlights the composition of the effective potential supporting the condensate wavefunction as its eigenstate with the corresponding eigenvalue $\mu$. It is clear that the radial flux that exists due to the spatially inhomogeneous phase of the condensate $\phi(r) \neq const$, forms an attractive potential that can trap a spatially localised bound state even in the absence of the external trapping $U_0=0$. In the linear limit (i.e. for very small condensates and/or nonlinearity) this effective trapping potential due to the flux is approximately given by:
\begin{equation}
 V_{J}(r) \approx \frac{\gamma^2_c}{4}\left(\bar{P}_0-1\right)^2r^2. 
 \label{effpot}
 \end{equation}
The effect of existence of the localised linear modes supported purely by the localised gain is known as "gain guiding" in optics \cite{siegman}.

\subsection{Numerical results}

\begin{figure}[here]
\includegraphics[width=8.5 cm]{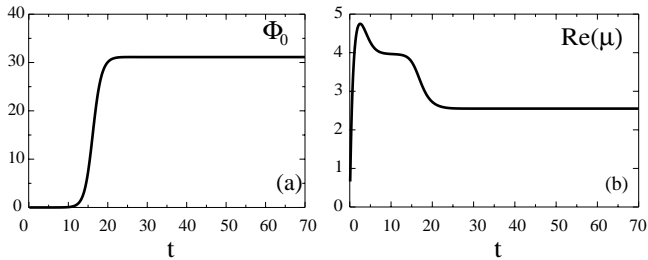} 
\caption{Typical time evolution of (a) the condensate peak density and (b) the chemical potential for $\bar{P}_0=3$, without a trapping potential.}
\label{dynamics_free}
\end{figure}

\begin{figure}[here]
\includegraphics[width=8 cm]{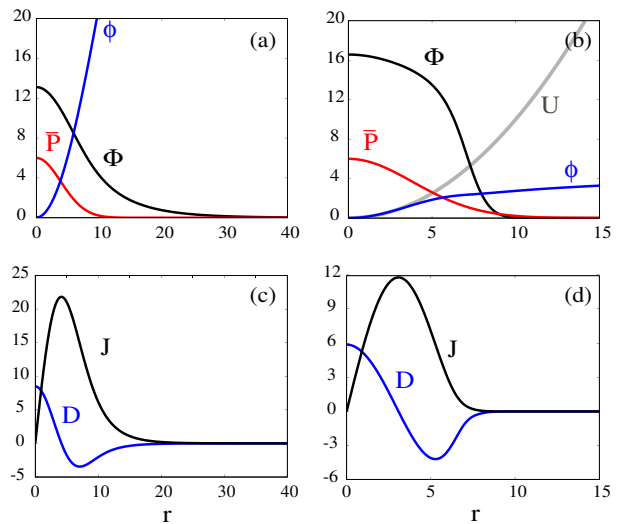} 
\caption{Typical radial shape of the condensate ground state in (a) trap-free and (b) harmonically trapped case. Both amplitude, $\Phi(r)$, and phase, $\phi(r)$, are shown in comparison to the Gaussian pump profile $P(r)/P_{th}$ with the amplitude $\bar{P}_0=P(0)/P_{th}=6$. Profiles of the radial flux $J$ and its divergence $D$ (scaled up by the factor of 10) for the ground state are shown for the (c) trap-free and (d) trapped cases, respectively. The harmonic potential of the depth $U^2_0=0.1$ is shown in (b) by the grey curve.}
\label{condensate}
\end{figure}

The existence of the localised steady state of the dissipative polariton BEC with the properties described in the previous section can be demonstrated by the numerical simulation of the time-dependent model equation (\ref{model}). Fig. \ref{dynamics_free} shows the typical dynamical behaviour and transition to the steady ground state ($m=0$) of the polariton component for intermediate values of pumping power ($\bar{P}_0=3$) in the trap-free regime. In addition, from the numerical simulations we can extract and plot the following dynamical quantity:
\begin{equation}
\mu(t)=-\frac{1}{4} \frac{\int_S \left[\nabla^2|\Psi|^2-2|\nabla\Psi|^2-4V|\Psi|^2\right] d\vec{r}}{\int_S |\Psi|^2d\vec{r}},
\end{equation}
where $S$ is the numerical integration domain. In the steady state limit this quantity coincides with the chemical potential of the condensate ${\rm Re}[\mu(t)] \to \mu$, ${\rm Im}[\mu(t)] \to 0$.
As can be seen from the dynamical simulations for $\bar{P}>1$ (Fig. \ref{dynamics_free}), the steady state regime is reached quickly, and the chemical potential is well defined. The condensate wavefunction displays characteristic inhomogeneous amplitude and phase profile $\phi(r)$ [seen in Fig. \ref{condensate}] with the limiting behaviour well described by Eqs (\ref{infty},\ref{zero}).

From Fig. \ref{condensate}(c) it is seen that the inhomogeneous phase of the condensate wave function results in the nonzero flux of the polariton superfluid. Due to the spatially localised pump, the polariton superfluid is continuously generated in the core area of the steady state $D>0$ and dissipated on its wings $D<0$. Thus the ground state of the strongly repulsive polariton BEC is a spatially localised {\em dissipative soliton}. Its spatial localisation is owed to the balance of the superfluid currents, and the internal structure of the currents is similar to that of a continuously "self-defocusing"  dissipative "antisoliton" of the generalised complex Ginzburg-Landau equation described in \cite{akhmediev}. 

The presence of a harmonic trap modifies the asymptotic behaviour of the condensate and phase according to Eq. (\ref{inftytrap}), as shown in Fig. \ref{condensate}(b). Nevertheless, the structure of the internal flow within the ground state of the condensate remains qualitatively the same, as seen in Fig.  \ref{condensate}(d).  In the trapped regime the internal flow of the polariton superfluid currents lead to significant distortions in the condensate density profile, $n_c(r)$. In particular, a wide pump with $\bar{P}_0\gg1$ leads to a well pronounced central peak [shown in Fig. \ref{3d} (b) for $\sigma^2=40$ and $\bar{P}_0=10$] noted in numerical simulations of a similar model of polariton BEC \cite{Berloff08} and also hinted at in experimental observations of a trapped exciton-polariton condensate \cite{Balili07}. 

\section{Trapped vs trap-free regime}

\begin{figure}[here]
\includegraphics[width=7.5 cm]{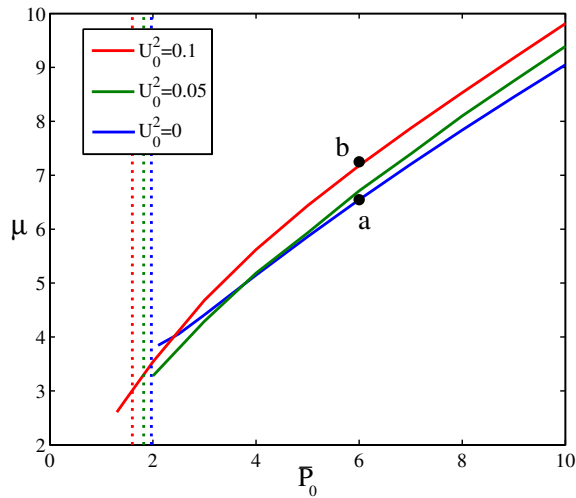} 
\caption{Dependence $\mu(\bar{P}_0)$ for the ground state $m=0$ without the harmonic potential (blue line) and in the presence of a harmonic trapping potential of the strength $U(r)=U^2_0r^2$. Vertical dashed lines indicate the cut-off value of $\bar{P}_0$ for the ground state given by the linear limit (\ref{linear}). Marked points (a) and (b) correspond to the condensate profiles shown in Fig. \ref{condensate}. The width of the pump spot is $\sigma^2=30$ in all cases.}
\label{mu}
\end{figure}

The ground $m=0$ state of the polariton condensate can be characterised by the dependence of chemical potential, $\mu$ on the parameters of the pump. In the absence of the trapping potential, this dependence is determined by the flux balance condition, and in the limit of low condensate densities the dependence of $\mu$ on the pump intensity is linear $\mu\sim \bar{P}_0=P_0/P_{th}$. In the presence of the harmonic potential, this condition is modified by the trap, as seen in Fig. \ref{mu}. 

For a fixed width of the pump, the cut-off value of $\bar{P}_0$ corresponds to the linear limit $|\psi|\to 0$, and deviates quite significantly from the value $\bar{P}_0=1$ determined by the threshold behaviour (\ref{th}) in the homogeneous excitation case. The linear limit is determined by the cut-off for the ground state in the effective 2D potential formed by the combination of the external harmonic trap $U(r)$, repulsive potential due to the spatially localised pump $V_R(r)=g_Rn^0_R$, and the internal inhomogeneous flux $V_J(r)$ given by Eq. (\ref{effpot}). This cut-off can be estimated from the condition that the effective potential in the linear limit is trapping (attractive) rather than anti-trapping (repulsive). Thus, the bound state appears in the effective potential at the value of $\bar{P}_0$ given by the positive root of the equation:
\begin{equation}
\bar{P}^2_0-2\left(1+\frac{4g_R}{\gamma_c R}\frac{1}{\sigma^2}\right ) \bar{P}_0 +\left(1+\frac{4U_0}{\gamma^2_c}\right)=0. 
\label{linear}
\end{equation}
According to this formula, for the parameters in Fig. \ref{mu}, the ground state appears in the trap-free system at $\bar{P}_0 \approx 1.97$. The presence of the harmonic trap with $U^2_0=0.1$ lowers the threshold to $\bar{P}_0 \approx 1.61$. The predicted tendency of the harmonic confinement to {\em lower} the threshold of the steady state formation compared to the trap-free case, agrees with the numerically calculated cut-off values in Fig. \ref{mu}. Above the threshold, the harmonic trap leads, for the same parameters of the pump, to the formation of the  steady state BEC with stronger spatial localisation and higher peak density, as can be seen from comparison of Figs. \ref{condensate}(a) and (b). 

\begin{figure}[here]
\includegraphics[width=7.5 cm]{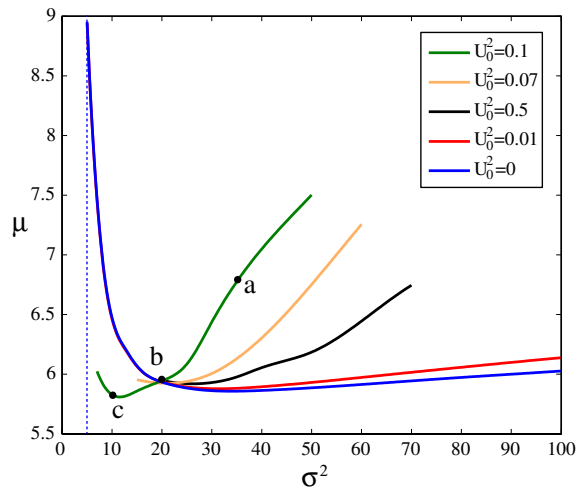} 
\caption{Dependence $\mu(\sigma^2)$ for the ground state $m=0$ without and with a harmonic trapping potential $U(r)=U^2_0r^2$ for $\bar{P}_0=5$. The vertical dotted line marks the cut-off value of the pump width for the trap-free ground state given by the linear limit expression (\ref{linear}) resolved with respect to $\sigma^2$. Marked points (a,b,c) correspond to the condensate profiles shown in Fig. \ref{profiles_vs_sigma}. }
\label{mu_vs_sigma}
\end{figure}

\begin{figure}[here]
\includegraphics[width=8.5 cm]{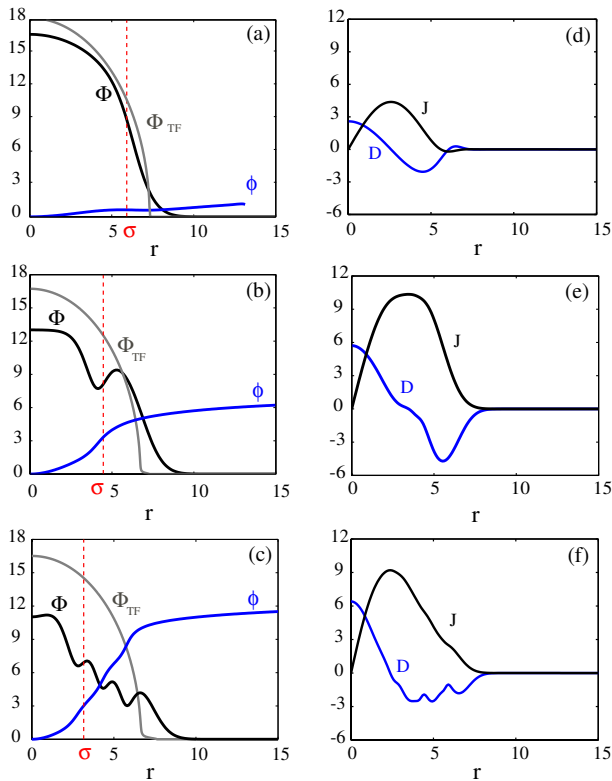} 
\caption{Radial shapes of the condensate ground states in the harmonically trapped case for $U^2_0=0.1$, $\bar{P}_0=5$ and the width of the pump spot $\sigma^2$ corresponding to the points marked in Fig. \ref{mu_vs_sigma}. (a-c) Both amplitude, $\Phi(r)$, and phase, $\phi(r)$, are shown in comparison to the condensate profile obtained in the Thomas-Fermi  approximation, $\Phi_{TF}$ (grey). (d-f) Profiles of the radial flux $J$ and its divergence $D$ (scaled up by the factor of 10) for the ground states shown in the panels (a-c).}
\label{profiles_vs_sigma}
\end{figure}

For a fixed strength of the harmonic confinement, the width of the pump spot determines different trapping regimes. Intuitively, for an excitation spot much wider than the trap, the limiting behaviour should be described by a homogeneous pump approximation $P(r)\approx P_0$ \cite{Wouters08,Berloff08}. In the opposite regime of a narrow excitation spot, the limiting behaviour should be close to that of a free condensate. Contrary to this intuitive picture, the dependence of the chemical potential $\mu$ on the width of the excitation spot, $\sigma^2$, calculated numerically shows dramatic differences for trapped and free condensates (Fig. \ref{mu_vs_sigma}), and the condensate density profiles are strongly modulated (see Fig. \ref{profiles_vs_sigma}). This is due to the presence of two competing spatial scales. One of them, $\sim \sigma$, is defined by the pump, and the other one $\sim r_{TF}$, is defined by the trapping potential and is given by the characteristic radius of the wave function in the Thomas-Fermi limit. The latter is obtained by neglecting both phase and density gradients in Eqs. (\ref{fluid}), as the solution to the equation:
\begin{equation}
U^2_0r^2_{TF}+\frac{\gamma_cg_R}{R}\bar{P}_0\exp\left(-\frac{r^2_{TF}}{\sigma^2}\right)=\mu.
\label{TF}
\end{equation}
In the regime of a wide pump, the condensate density can be reasonably well approximated by the radially-symmetric Thomas-Fermi profile with  the radius $r_{TF}$  given by Eq.(\ref{TF}), as can be seen in Fig.  \ref{profiles_vs_sigma}(a). In this regime, the ground state in the form of a dissipative soliton becomes dynamically unstable with respect to azimuthal perturbations for $\bar{P}_0\gg1$, as established in \cite{Berloff08}. There it was shown that this instability can lead to rotational symmetry breaking whereby multiple vortex states enter the condensate. 

For a very narrow pump, $\sigma^2\ll r^2_{TF}$, the Thomas-Fermi radius exactly coincides with the well-known expression for a conservative BEC with repulsive inter-particle interactions in a harmonic trapping potential: $r^0_{TF}=\sqrt{\mu/U^2_0}$. In our numerical simulations we find that $r_{TF}\approx r^0_{TF}$ for a wide range of values $\sigma<r_{TF}$. In this parameter regime the condensate density profile strongly deviates from the Thomas-Fermi profile [Figs. \ref{profiles_vs_sigma}(b,c)], as neither phase nor density gradients can be neglected. The condensate experiences strong modulations of density with a notable density dip along the line $D=0$ separating the regions of loss and gain.  The condensate peak density is also damped as the pump narrows down and the dip at $r=0$ appears for the values of $\sigma$ corresponding to the negative value of  $q_{-}$ in (\ref{zero}). We note that the point (b) in Fig. \ref{mu_vs_sigma} corresponds to $q_{-}=0$ for $U^2_0= 0.1$. Under the conditions of a strong ($\bar{P}_0\gg1$) narrow pump the steady state is dynamically unstable and exhibits strong peak density oscillations. 

\begin{figure}[here]
\includegraphics[width=8.5 cm]{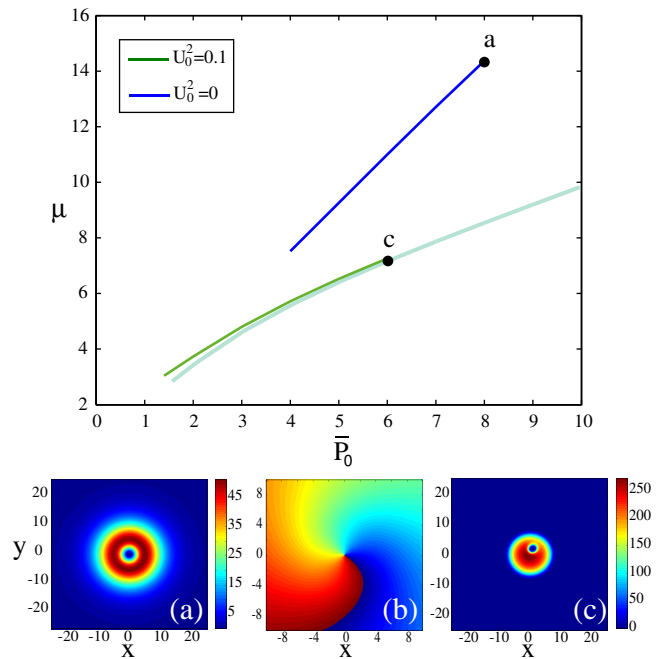}
\caption{Top: Dependence $\mu(\bar{P}_0)$ for the vortex state $m=1$ without and with the harmonic trapping potential $U(r)=U^2_0r^2$ for $\sigma^2=30$. Semi-transparent line indicates the corresponding dependence for a trapped $m=0$ state from Fig. \ref{mu}. Bottom: (a) Condensate density and (b) phase for a dynamically unstable $m=1$ stationary vortex state in a trap-free condensate at $\bar{P}_0=8$. (c) Metastable rotating vortex state supported by a harmonic trap ($U^2_0=0.1$) at $\bar{P}_0=6$}
\label{mu_vortex}
\end{figure}

\section{Single vortex state}

Spatially localised dissipative vortices are similarly found as higher-order steady states of the system with $m\neq 0$ \cite{tristram}, both in a trap-free and harmonically trapped polariton BEC. Numerically, the system relaxes to a single steady state vortex when a phase factor $\exp(im\theta)$ is imprinted onto a condensate "seed" at the initial stages of evolution. The dependence of the chemical potential $\mu$ on the pump amplitude $\bar{P}_0$ is plotted in Fig. \ref{mu_vortex} for a  vortex with $m=1$. As expected, the chemical potential and the cutoff for this excited state is higher than those of the ground state of the trap-free condensate. In contrast, in a harmonic potential the vortex and the ground state become nearly degenerate (see Fig. \ref{mu_vortex}). 

The global phase of the vortex is imposed not only  by the vorticity but also by the nonzero flux, and therefore depends on both azimuthal ($\theta$) and radial ($r$) coordinates. Using the expression (\ref{zero}) for the radial dependence of the phase near the vortex core, we find that the lines of the constant phase follow a spiral trajectory described by the equations:
\begin{equation}
x_s(\theta)=\sqrt{\theta/q_+}\cos \theta, \quad y_s(\theta)=\sqrt{\theta/q_+}\sin \theta \nonumber
\end{equation} 
This spiral phase structure is shown in Fig. \ref{mu_vortex}(b) for the numerically found localised vortex state in Fig. \ref{mu_vortex}(a), without the trapping potential. It is a signature of the polariton BEC vortex observed in the experiments \cite{pvortex08}, and also appears as a characteristic feature of vortex states in other dissipative systems \cite{rosanov,2d_kartashov_vortex,spiral}.

The ground state ($m=0$) of the polariton condensate is dynamically stable in the absence of a trap. In contrast, we find that a single vortex created in a localised steady state BEC by phase imprinting is dynamically unstable and, in the absence of potential, spirals out of the condensate, as predicted for the case of trapped dissipative BEC of alkali atoms \cite{vortex_atom,dissipative_v}.
As can be seen from comparison of Figs. \ref{mu} and \ref{mu_vortex}, the addition of a stationary  (i.e. non-rotating) harmonic potential dramatically modifies the threshold of the optical pumping required to sustain a steady state of a polariton BEC with an angular momentum and leads to the possibility of a long-term survival of the vortex in the form of an eccentric rotating state shown in Fig. \ref{mu_vortex}(c).  The detailed study of the stabilising effect of the trapping potential on the vortex dynamics and the dynamics near the onset of rotational symmetry breaking \cite{Berloff08} is beyond the scope of this study. However, we have confirmed that a wide pump (corresponding to small density gradients in the harmonic trap) tends to stabilise the vortex state. 

\section{Conclusions}

We have characterised  the 2D stationary regime of the polariton BEC in the framework of an open-dissipative Gross-Pitaevskii model coupled to an exciton reservoir and described, both analytically and numerically, properties of the ground and excited states (vortices) depending on the pump and trapping parameters.  We have shown that the spatial localisation of the condensate, even in the absence of a trapping potential, is supported by the balance of the internal superfluid flows established by an inhomogeneous nonlinear gain due to optical pump and linear loss due to the decay of the polaritons. The ground state of the condensate can therefore be described as a {\em continuously self-defocusing dissipative soliton}. We have also investigated localised dissipative vortex states and have shown that they can display metastable behaviour in non-rotating trapping potentials.

Finally, we note that the open-dissipative model with the inhomogeneous pump used here and successfully employed for theoretical description of experiments  with microcavity polaritons, has many similarities to optical systems with localised  gain landscapes \cite{1d_kartashov_ol,2d_kartashov_vortex}. These and similar studies suggest that the 2D stationary regime of polariton BEC can potentially display a rich variety of localised states which are not exhausted by radially symmetric configurations. In particular, the experiments with polariton BEC excited by elliptical optical pump have demonstrated pattern formation due to the appearance of standing waves \cite{pattern_formation}. Furthermore,  the recent  experiment with polariton BECs created by two pump spots \cite{oscillations_nature} demonstrates novel possibilities arising from {\em interaction of two dissipative solitons with variable separation}. The clear understanding of the structure of phase gradients underpinning the existence of steady-state polariton condensates presented in our work enables us to gain an immediate insight into the properties and outcomes of such interactions. The possibility to create a rich variety of localised states by varying the spatial properties of the off-resonant optical pump are currently under investigation and will be reported elsewhere.

\end{document}